\documentclass[
 reprint,
superscriptaddress,
 amsmath,amssymb,
 aps,
prb,
]{revtex4-2}

\usepackage{graphicx}
\usepackage{newtxtext,newtxmath}
\usepackage{float}
\usepackage{hyperref}
\usepackage{physics}

\usepackage{xr}
\usepackage{lastpage}
\renewcommand{\thesection}{\arabic{section}}
\renewcommand{\thesubsection}{\thesection.\arabic{subsection}}
\renewcommand{\thesubsubsection}{\thesubsection.\arabic{subsubsection}}

\newcommand{\kopt}{k_\text{opt}}

\newcommand{\fO}{f_0}
\newcommand{\df}{\Delta f}
\newcommand{\Fopt}{F_\text{opt}}
\newcommand{\Qeff}{Q_\text{eff}}
\newcommand{\QO}{Q_0}
\newcommand{\geff}{\gamma_\text{eff}}
\newcommand{\gO}{\gamma_0}
\newcommand{\kc}{k_c}
\newcommand{\ktherm}{k_\text{therm}}
\newcommand{\Pcirc}{P_\text{circ}}
\newcommand{\Ctherm}{C_\text{therm}}
\newcommand{\keff}{k_\text{eff}}
\newcommand{\dw}{\Delta \omega}
\newcommand{\weff}{\omega_\text{eff}}
\newcommand{\wO}{\omega_0}

\newcommand{\Icirc}{I_\text{circ}}
\newcommand{\Iinc}{I_\text{inc}}
\newcommand{\Irefl}{I_\text{refl}}
\newcommand{\Acirc}{A_\mathrm{circ}}

\begin{document}
\date{\today}

\title{Control of quality factor of atomic force microscopy cantilever by cavity optomechanical effect}

\author{Noah Austin-Bingamon}
\author{Binod D.C.}
\affiliation{Materials Science, Engineering and Commericialization program,
Texas State University, San Marcos, Texas, 78640, USA}
\author{Yoichi Miyahara}
\email{yoichi.miyahara@txstate.edu}
\affiliation{Materials Science, Engineering and Commericialization program,
Texas State University, San Marcos, Texas, 78640, USA}
\affiliation{Department of Physics, Texas State University, San Marcos, Texas, 78640, USA}

\begin{abstract}
Quality factor plays a fundamental role in dynamic mode atomic force microscopy.
We present a technique to modify the quality factor of an atomic force microscopy cantilever within a Fabry-P\'erot optical interferometer.
The experimental setup uses two separate laser sources to detect and excite the oscillation of the cantilever. 
While the intensity modulation of the excitation laser drives the oscillation of the cantilever, 
the average intensity can be used to modify the quality factor via optomechanical force, without changing the fiber-cantilever cavity length. 
The technique enables users to optimize the quality factor for different types of measurements
without influencing the deflection measurement sensitivity.
An unexpected frequency shift was also observed and modelled as temperature dependence of the cantilever's Young's modulus, which was validated using finite element simulation.
The model was used to compensate for the thermal frequency shift. The simulation provided relations between optical power, temperature, and frequency shift.

\end{abstract}

\maketitle
\section{Introduction}
The mechanical quality factor ($Q$ factor) of an atomic force microscopy (AFM) cantilever is one of the most important 
quantities which determine the measurement noise and measurement speed
\cite{albrechtFrequencyModulationDetection1991b}.
A number of techniques for increasing or decreasing  the $Q$ factor of AFM cantilevers have been reported \cite{millerEffectiveQualityFactor2018b}. 
In these techniques, an external damping force is applied to feed energy into or out of the resonator,
which modifies the effective $Q$ factor.
The type of actuation force can be optical \cite{mertzRegulationMicrocantileverResponse1993b}, capacitive \cite{huefnerMicrocantileverControlCapacitive2012a}, and mechanical forces \cite{anczykowskiAnalysisInteractionMechanisms1998b}.
The means to produce the damping force can be 
an active feedback of the displacement (deflection) signal with an external circuit \cite{mertzRegulationMicrocantileverResponse1993b,anczykowskiAnalysisInteractionMechanisms1998b, weldFeedbackControlCharacterization2006, poggioFeedbackCoolingCantilever2007a,huefnerMicrocantileverControlCapacitive2012a}
and a coupling the cantilever to an environment which produces a position (deflection)-dependent force such as an optical cavity 
(cavity optomechanical force)
\cite{metzgerCavityCoolingMicrolever2004,
metzgerOpticalSelfCooling2008, 
holscherEffectiveQualityFactor2009a,
Troger2010, vonschmidsfeldControllingOptomechanicsCantilever2015}.
The optical force (radiation pressure and photothermal) acting on the cantilever is proportional to the optical intensity on the cantilever surface, 
which is dependent on the position (deflection) of the cantilever in the optical cavity due to the periodic modulation of the optical intensity caused by optical interference.
The resulting position dependent optical force acts as an optical spring 
whose sign depends on the slope of the interferogram (optical intensity vs position relation). 
When there is a time delay between the position and the optical force,
the quadrature component of the optical force acts as a damping force,
leading to the change in effective $Q$ factor \cite{metzgerOpticalSelfCooling2008, miyaharaDissipationModulatedKelvin2018}.

In the fiber-optic interferometer setup,
it is well-known that the effective $Q$ factor of a cantilever can be modified by changing the cavity length (fiber-cantilever distance) and the laser power 
\cite{metzgerCavityCoolingMicrolever2004, 
metzgerOpticalSelfCooling2008,
holscherEffectiveQualityFactor2009a, 
Troger2010, 
vonschmidsfeldControllingOptomechanicsCantilever2015, 
millerEffectiveQualityFactor2018b}.
In the practical operation of the AFM, however, these two parameters are chosen such that the cantilever deflection measurement sensitivity is optimized.
Therefore, with the conventional interferometer setup with only one laser,  one cannot change these two parameters to modify the effective $Q$ factor. 
Fiber-optic interferometer setups with two lasers have been reported \cite{weldFeedbackControlCharacterization2006, miyaharaOpticalExcitationAtomic2020d}, 
in which one laser (detection laser) is used to detect the cantilever deflection 
and another laser (excitation laser) is used to excite the cantilever oscillation.
Weld and Kapitulnik used the two laser setup for modifying the effective $Q$ factor of the AFM cantilever by an external feedback circuit 
\cite{weldFeedbackControlCharacterization2006}.  
Miyahara \textit{et al.} demonstrated that the two laser setup enables to obtain clean cantilever resonance spectra free from spurious resonance features 
\cite{miyaharaOpticalExcitationAtomic2020d}, which are critically important for accurate dissipation measurement 
\cite{labudaDecouplingConservativeDissipative2011}. 
In both studies, the intensity of the excitation laser is modulated at the resonant frequency of the cantilever to excite
the cantilever oscillation.
In this report, we show that a similar setup can be used to modify the effective $Q$ factor via cavity optomechanical force by changing the average intensity of the excitation laser, without changing the deflection measurement sensitivity. 

\section{Experimental}
\begin{figure}[H]
    \centering        
     \includegraphics[width=.48\textwidth]{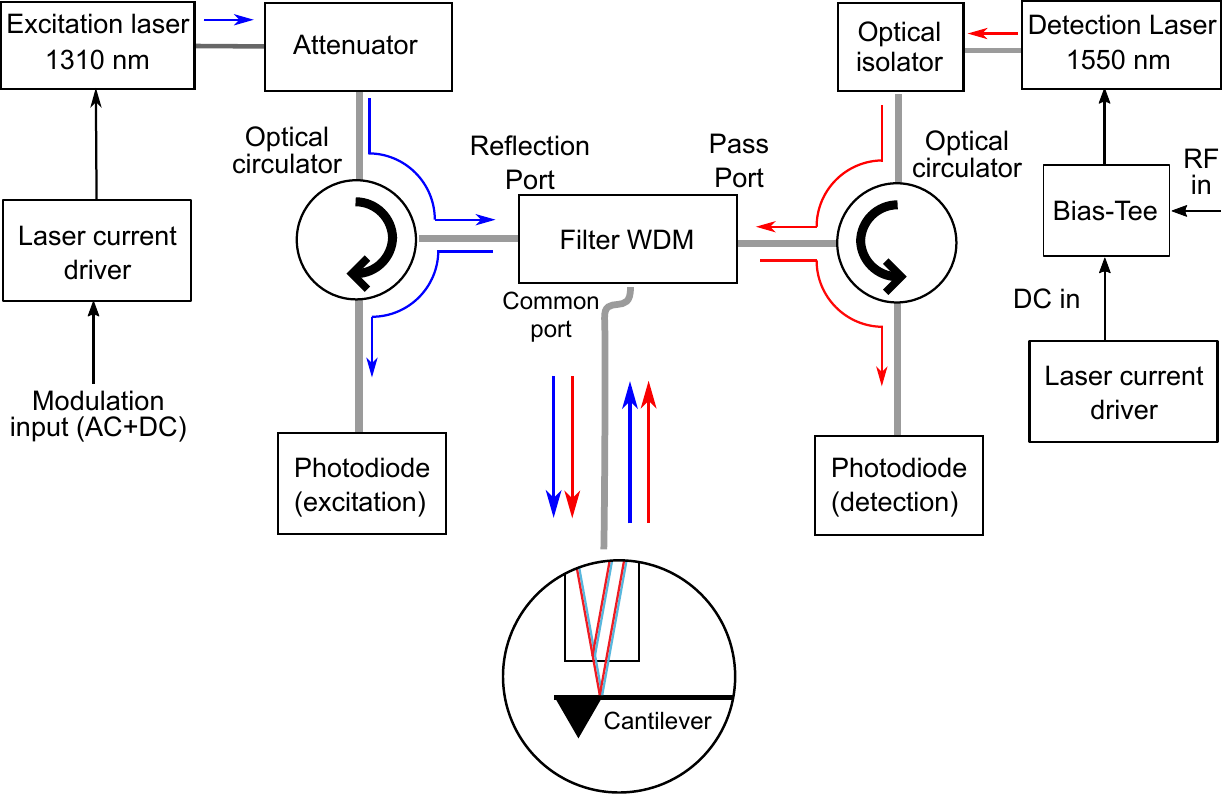
     }
    \caption{Schematic diagram of optical circuit with two laser diodes.}
    \label{fig:optical circuit}
\end{figure}

We performed the experiments with a home-built low-temperature atomic force microscope which is operated in a cryogen-free dilution refrigerator (LD250, Bluefors). 
All the experiments were performed at 4~K in cryogenic vacuum (pressure lower than $3\times 10^{-6}$~mbar) without running the dilution unit. 
Figure~\ref{fig:optical circuit} shows the schematic diagram of 
the optical setup, which is similar to what has been reported by Miyahara \textit{et al.}~\cite{miyaharaOpticalExcitationAtomic2020d}.
The setup utilizes two laser diodes at wavelengths of 1550~nm (PL15CE001FAG-A-1-01, NECSEL (Ushio)) 
and 1310~nm (PL13AG0021FAG-A-1-01, NECSEL (Ushio)). 
These wavelengths are commonly used optical communications wavelengths, which greatly reduces the cost of the apparatus. 
In the experiments reported here, the 1550~nm laser is used for interferometric detection of the cantilever oscillation, while the intensity of the 1310~nm laser is modulated at the cantilever resonance frequency to excite the cantilever via photothermal force and radiation pressure.
%This choice of the wavelength is arbitrary. 
We selected the 1310~nm laser (excitation laser) for the excitation because only the 1550~nm laser (detection laser) was equipped with an optical isolator in the original setup. 

If the 1310~nm laser is equipped with an optical isolator, using the 1310~nm laser for detection would give a better deflection measurement sensitivity because the sensitivity is inversely proportional to the wavelength.
The intensity of the excitation laser is sinusoidally modulated via the drive current using a laser current source with an external modulation input (LDC202C, Thorlabs).
The bandwidth of the modulation  of the current source ranges from DC to 250~kHz ($-3$~dB) and the modulation sensitivity is 20~mA/V.
The optical outputs from both lasers are multiplexed to a single optical fiber using a filter wavelength division multiplexer (FWDM, FWDM-1513, AFW Technology). 
Reflected light returning from the fiber end is diverted to InGaAs photodiodes (PDINAS0701FAD-A-0-01, NECSEL) using optical circulators (CIR-3-13-L-1-2 and CIR-3-15-L-1-2, AFW Technology). 
Therefore the reflected light is separated by wavelength and separated from the outgoing light, which makes readout of the reflected power straightforward. 
All the connectors for the optical fibers are of FC/APC (angled polished face) type to minimize optical back-reflection. 
The photodiodes are connected to custom-built transimpedance amplifiers to convert the photocurrents to voltage signals.

The 1550~nm (detection) laser diode has an integrated optical isolator, which prevents stray reflection within the optics from reentering the laser diode. 
This helps to improve the stability of the diode and reduce noise. 
To further reduce noise from stray reflections, the 1550~nm laser diode is modulated at around 10~MHz using an RF source and a bias tee (PBTC-1GW, Mini-Circuit), which reduces the coherence length of the laser light.
Because the fiber cantilever gap is very small (several $\mu$m) in the normal AFM operation, a long coherence length is not necessary to maintain interference. 
However, the reflections within the fiber optics, such as at connectors, can produce unwanted interferometric effects which act as pickups of mechanical vibration and thermal expansion of the optical components. 
Because the lengths of these unwanted interferometers are much longer, on the order of the lengths of the fiber segments, a reduced coherence length effectively eliminates interference here.

We coated the cleaved end of the optical fiber with titanium dioxide film by dip-coating technique 
 to enhance reflectivity of the fiber-vacuum interface 
 \cite{subba-raoImprovingHighresolutionFiberoptic2009b}. 
We typically achieve the reflectivity of about 30~\%.
The gap between the fiber and the cantilever surface forms a low-finesse Fabry-P\'erot (FP) cavity. 
The fiber-cantilever gap (cavity length) is adjustable using a piezoelectric stick slip motor whose step size can be as small as $\sim$10~nm.
%(Fig.~\ref{fig:fig:fiber_walker}).
Adjusting the fiber-cantilever gap yields an interferogram (interference fringe) for each laser. 
To maximize the sensitivity for deflection measurement, 
we set the fiber position on a high slope region of the interferogram 
 for the detection laser (1550~nm). 
The slope determines the detection sensitivity as well as the optical spring force (cavity optomechanical force) due to the detection laser.
In the conventional single laser setup, the optical spring force caused by the detection laser is responsible for the observed changes in effective $Q$ factor.
Therefore, controlling the effective $Q$ factor necessitates changing the slope either by changing the cavity length or the optical power.

The dual-laser scheme has the following advantage compared to the optical cantilever excitation scheme with only a single laser \cite{celikRadiationPressureExcitation2017b}
and is critical to the effective $Q$ factor modulation technique presented here.
Because the detection (1550~nm) laser is dedicated solely to detection, it does not need to be modulated for the cantilever excitation. 
The AC component of the reflected detection laser power directly reflects the cantilever oscillation, and there is no need to subtract the modulation signal unlike in a single laser setup
where a single laser is used for both detection and excitation \cite{celikRadiationPressureExcitation2017b}. 
Additionally, we can vary the average (DC) intensity of the excitation (1310~nm) laser by varying the DC component of the modulation signal which is applied to the laser current driver.
This permits us to vary the optomechanical spring force which is caused by the excitation (1310~nm) laser without changing the deflection measurement sensitivity.
A lock-in amplifier (MFLI, Zurich Instruments) was used to drive the cantilever oscillation and capture and demodulate the deflection signal.
The output of the lock-in amplifier is connected to the modulation input of the current driver for the excitation laser. 
Adding a DC offset voltage to the lock-in output enables to change the DC power of the excitation laser, which is exploited to modify the effective $Q$ factor.

Measurements of the $Q$ factor modulation were taken at several fiber positions using an automated program written with Python and leveraging the LabOne Python API from Zurich instruments. 
The fiber position was controlled by using a MK3-PLL Signal Ranger with open-source GXSM control software \cite{zahlOpenSourceScanning2010}. 
The GXSM software was controlled remotely from Python via a custom socket server interface written in Python. The fiber is stepped by providing a sawtooth burst from the Signal Ranger to the stick-slip motor's amplifier. 

A control sequence was implemented in the main Python script to step the fiber over many positions. 
At each position, the DC current of the excitation laser was swept over 5 values ranging from 10~mA to 42~mA (corresponding to a power range of 67~$\mu$W to 2023~$\mu$W at the coated fiber end). 
For each DC power, a frequency sweep was performed from which the $Q$ factor was obtained by a Lorentzian fit. 
The relevant experimental parameters are summarized in Table~\ref{table}. 
\onecolumngrid

\begin{table}[h!]
    \centering
    \begin{tabular}{lc}
        \hline  \hline
        Parameter & Value\\
        \hline 
        Cantilever resonant frequency, $\fO$ & 143~kHz\\
        Cantilever spring constant, $\kc$  &  20~N/m\\
        Cantilever coating & Pt, both front and back\\
        1550~nm (detection) laser power at coated fiber end & 670~$\mu$W\\
        1310~nm (excitation) laser power at coated fiber end & up to 2023~$\mu$W\\
        Measurement temperature & 4~K\\
        \hline \hline
    \end{tabular}
    \caption{List of experimental parameters}
    \label{table}
\end{table}

\twocolumngrid
\section{Results and Discussion}
\subsection{Cavity length and excitation DC power dependence of the $Q$ factor and the resonance frequency}
\begin{figure}
\centering
\includegraphics[width=0.99\textwidth]{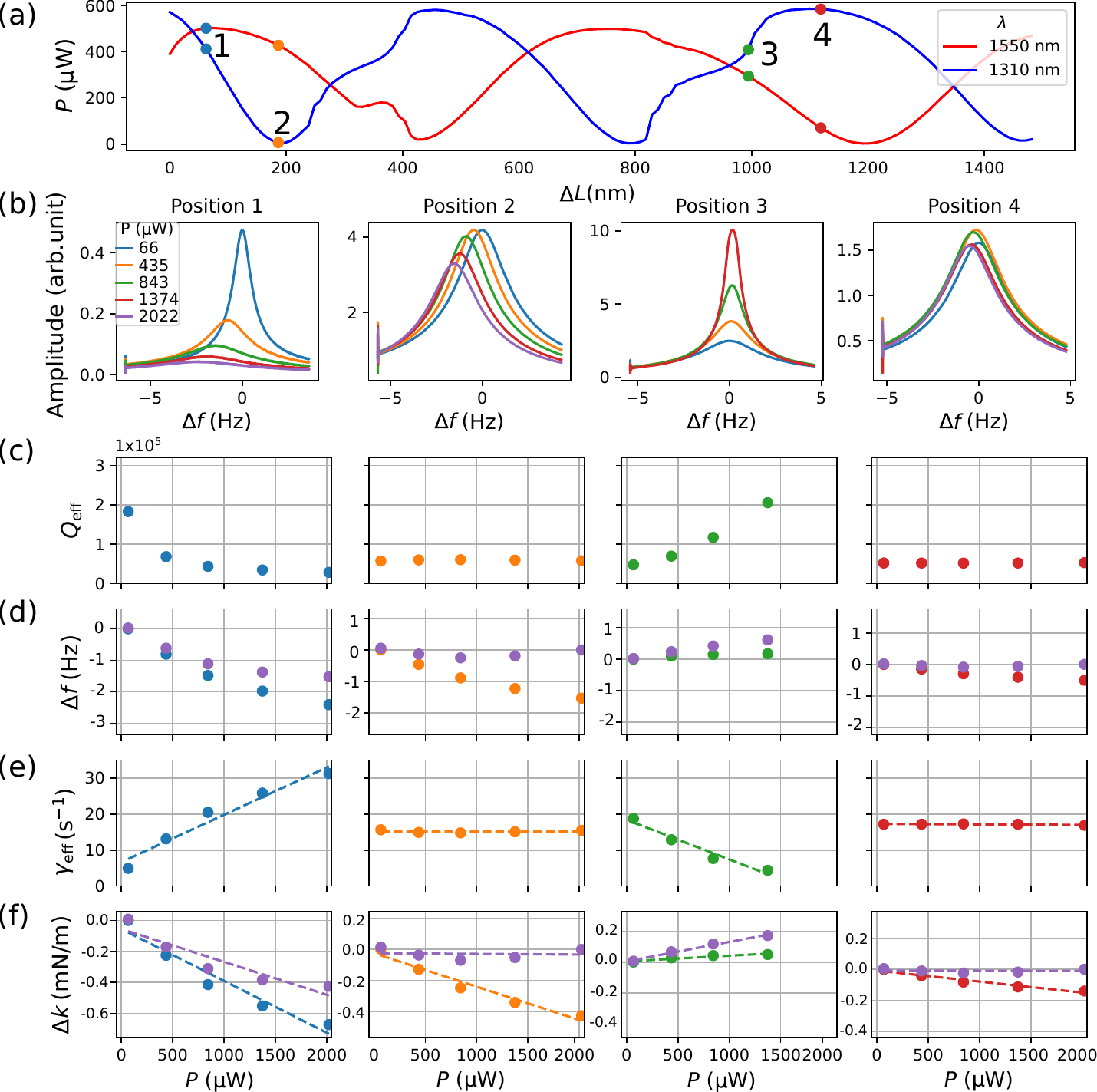}
\caption{\label{fig:DC Sweeps B K Q f0 at p 4 6 19 21} 
(a) Interferograms of detection laser (red) and excitation laser (blue). 
The numbers 1, 2, 3, 4 indicate the measurement positions.
(b) Average excitation laser power, $P$, dependence of the resonance curves measured at the positions 1, 2, 3, 4. Frequency is shown relative to the resonant frequency at the lowest laser power.
(c) Effective $Q$ factor, (d) Resonant frequency shift, shown relative the resonant frequency at the lowest laser power, $\df$, 
(e) Effective damping factor $\geff$, and 
(f) Optically induced spring constant versus average excitation laser power, $P$. 
In (d) and (f), the corrected values which exclude the thermally induced effect are shown as violet markers.
In (e) and (f), dashed linear lines are the least square linear fit of the experimental plots.}
\end{figure}

Figure~\ref{fig:DC Sweeps B K Q f0 at p 4 6 19 21} shows the effective $Q$ factor $\Qeff$, resonant frequency shift $\df$, effective damping $\geff$, and optical spring constant $\Delta k$,  as a function of the cavity length, $\Delta L$, and the DC optical power of the excitation laser, $P$.
Figure~\ref{fig:DC Sweeps B K Q f0 at p 4 6 19 21}a shows the interferograms of both detection and excitation lasers.

The cavity lengths (fiber positions) at which measurements were performed are indicated by numbers (1-4) and colored markers.
At each measurement position, we measured the frequency response of the cantilever at five different DC power settings of the excitation laser. 
Figure~\ref{fig:DC Sweeps B K Q f0 at p 4 6 19 21}b shows the measured frequency responses.
Figure~\ref{fig:DC Sweeps B K Q f0 at p 4 6 19 21}c and d
show the extracted effective $Q$ factor, $\Qeff$, and the shift of resonant frequency, $\Delta f$ versus the excitation laser power, $P$, respectively. 
Figure~\ref{fig:DC Sweeps B K Q f0 at p 4 6 19 21}e and f show the effective damping, $\geff$, and optical spring constant, $\Delta k$ which are extracted from $\Qeff$ and $\Delta f$, respectively.
Here we use the relations,
$\geff = 2\pi (\fO + \df)/\Qeff \approx 2\pi \fO/\Qeff$, and
$\Delta k = 2 \kc \times \Delta f/\fO$
where $\fO$ is the resonant frequency measured at the lowest power.
\onecolumngrid

\twocolumngrid
Positions 1 and 3 represent the locations with opposite slope on the excitation (1310~nm) laser fringe (blue line). 
Position 1 is on a downward slope, where a negative optical spring constant is expected.  
We observe a significant decrease in the effective $Q$ factor from 183,000 to 29,000 with increasing the excitation laser DC power.
The resonant frequency decreases by 2.3~Hz.
Position 3 is on an upward slope, where a positive optical spring constant is expected. 
A comparable increase in the $Q$ factor from 48,000 to 206,000 is observed with increasing the excitation laser DC power, as would be expected from a time delayed positive spring force.
However, the increase in the frequency shift is much smaller ($-0.17$~Hz) than that observed at position 1.
We ascribe this unexpected asymmetric change in the frequency shift to the temperature dependence of the cantilever's Young's modulus,
causing an additional negative frequency shift 
(thermal frequency shift) with increasing optical power.

At positions 2 and 4,
the effective $Q$ factor is almost constant
over the whole range of the excitation laser DC power.
This behavior is consistent with the nearly zero slope of the excitation laser fringe at these positions.
The frequency shift, however, show an appreciable decrease at both positions which is inconsistent with the nearly zero slope.
These unexpected negative frequency shifts can also be explained by the thermal frequency shift.
We will discuss the thermal frequency shift in more detail in the next section.

\subsection{Effect of optical absorption on resonance frequency shift}
We propose a model to explain the unexpected frequency shift based on heating by optical absorption and the temperature dependence of the cantilever's Young's modulus.
The Young's modulus in turn affects the spring constant and resonant frequency of the cantilever.
Kazantsev \textit{et al.} \cite{kazantsevCantileverTemperatureCharacterization2006} observed a significant frequency shift in an AFM cantilever due to thermal coupling to the sample, with sample temperatures ranging from 15 to 399~K. The effect was ascribed to a change in the Young's modulus. 
Likewise Boyd \textit{et al.} found a 0.5~\% change in the Young's modulus and a decrease in resonant frequency with temperature over the range of 200 to 290~K. 
From Boyd \textit{et al.} we derived a linear temperature coefficient for the Young's modulus of $-5.55 \times 10^{-5}$~(K$^{-1}$) \cite{boydMeasurementTemperatureCoefficient2013a}.
A similar interplay between the thermally induced and optomechanically induced resonant frequency shifts was reported by Vy \textit{et al.} 
\cite{vyCancelationThermallyInduced2016c}.

To assess whether heating by our interferometer is sufficient to account for the unexpected frequency shift, we performed a finite element eigenfrequency study with COMSOL Multiphysics. 
We modelled our cantilever using a temperature dependent function for the Young's modulus and performed thermal studies to predict the temperature of the cantilever under varying optical power.
The details of the study are discussed below.

We modelled a rectangular silicon cantilever with dimensions of 
$240~\text{(length)} \times 20~\text{(width)} \times 7~(\text{thickness})~\mu$m with a 25~nm thick platinum coating on both tip and backside surfaces. 
The material models from the COMSOL material library include temperature dependent functions for thermal conductivity and heat capacity for both silicon and platinum.
For silicon, we defined the temperature dependent Young's modulus as follows.
Deviation from the reference temperature (4~K) produced a linearly proportional deviation in the Young's modulus via the linear temperature coefficient from Ref.~\cite{boydMeasurementTemperatureCoefficient2013a}. 
The initial Young's modulus at 4k was given a value of 170~GPa which is the default in COMSOL's built in silicon model. 
This value is not too critical, since we are interested in the proportional change in Young's modulus.
The cantilever base was fixed at a temperature of 4.00~K, representing the temperature of the AFM body.

The COMSOL simulation was performed in two steps. First a thermal study under various heat flux inputs was performed to obtain the temperature profile of the cantilever. 
Heat flux was applied to the cantilever in a circular region near the free end, representing the region in proximity to the optical fiber core. 
The magnitude of the heat flux was determined from the reflectance of the platinum coating and the incident optical power. 
We used a surface reflectance of 0.8 at the platinum cantilever coating. 
This is a conservative value for both the 1550~nm and 1310~nm wavelengths derived from Ref.~\cite{coblentzReflectingPowerVarious1910a}.
The incident powers used in the simulation were chosen to be comparable to those in the experiment. In the simulation, we chose to ignore any effect of cavity enhancement. In our low finesse cavity the enhancement is on the order of 1, so the power exiting the fiber end is representative of the range of powers seen by the cantilever.
Laser powers exiting the fiber end in the experiment were a constant 670~$\mu$W for 1550~nm and a range from 67 to 2022~$\mu$W for 1310~nm.
Using the reflectance value from Coblentz, these powers result in an absorbed heat flux ranging from 147 to 539~$\mu$W with temperature at the cantilever tip ranging from 4.52 to 5.45~K. 
The temperature profile is shown for the highest applied power in Supporting figure Fig.~S1.

The temperature profiles from the thermal study were next used as inputs to the mechanical model to calculate the temperature dependent Young's modulus. 
Eigenfrequencies were calculated for each value of the input heat flux. 
The study was run with and without modelling the effect of thermal expansion on the eigenfrequencies. The effect of thermal expansion was insignificant compared to that of the Young's modulus, in agreement with Kazantsev \textit{et al.}. 
Finally, in a separate study, heat flux ranging from $0\sim 1600~\mu$W was introduced to the cantilever, resulting in a larger temperature rise of $\sim 2.1$~K at the cantilever tip (see Supporting figure Fig.~S1 and Supporting figure Fig.~S2.

\begin{figure}
\centering
\includegraphics[width=.48\textwidth]{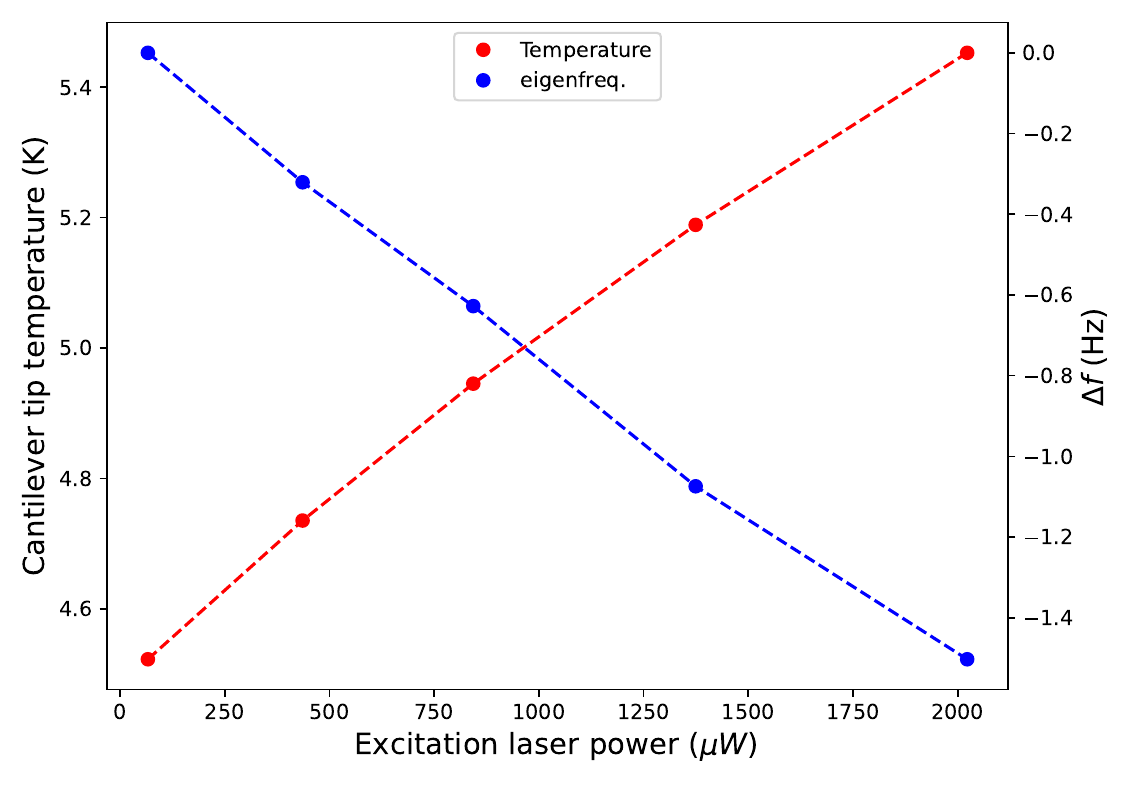}
\caption{\label{fig:cantilever eigs and temps comsol} Simulated cantilever tip temperature (red) and eigenfrequency (blue) of the cantilever versus excitation laser power.}
\end{figure}

Figure~\ref{fig:cantilever eigs and temps comsol} shows the simulated cantilever tip temperature (red) 
and the resonant frequency shift (blue)
as a function of the excitation laser power.
The simulated frequency shift of $-1.50$~Hz at the maximum applied power is in remarkable agreement with our results. 
Position 2 in Fig.~\ref{fig:DC Sweeps B K Q f0 at p 4 6 19 21}, where no optomechanical spring force is expected, displays a frequency shift of $-1.58$~Hz. 

The COMSOL simulation result shows that this is a feasible model. 
We applied the principle to our data by splitting our optically induced spring constant shift into an optomechanical component and a thermal component, as $\Delta k = \kopt + \ktherm$. 
The optomechanical spring constant term, $\kopt$, is proportional to the gradient of the optical power within the cavity, while the thermally induced change in spring constant, 
$\ktherm$, is dependent on the magnitude of the circulating power within the cavity. 
The circulating power was derived from a fitting of the excitation laser interferogram \cite{ismailFabryPerotResonatorSpectral2016}
(See more details about the fitting in Supporting information~S2). 
The fitting yielded the reflectivity at the fiber-vacuum interface, $R_1 = 0.251$, and the reflectivity at the cantilever surface, $R_2 = 0.155$. 
While the fitted reflectivity at the fiber-vacuum interface agrees reasonably well with the expected value of $\sim0.3$,
the fitted reflectivity at the cantilever surface is much lower than the value (0.8) given by Coblenz. 
This is likely due to coupling losses from the cavity due to imperfect alignment of the mirrors \cite{wilkinsonAnalyticalModelLow2011a}.
The fitted cavity reflectivities were used to calculate the circulating power as a function of the cavity length and applied optical power (Supporting figure Fig.~S5). 

Finally the calculated circulating power was used to predict the thermal spring constant shift via a linear coefficient, $\Ctherm$, such as
$\ktherm = \Ctherm \Pcirc$.
To find the value of $\Ctherm$ we leveraged the zero slope fringe positions (positions 2 and 4). 
At these positions we expect no change in frequency shift and no $\Delta k$ due to the changing laser power
if only the cavity optomechanical effect were present. 
We sought a value of $\Ctherm$ which nullifies the predicted $\Delta k$ at both of the zero slope locations when subtracting the predicted $\ktherm$ from the measured $\Delta k$.
To achieve the desired cancellation at the zero slope positions, we found that a vertical shift needed to be applied to the circulating power curves (see Supporting figure Fig.~S6). 
A negative shift equal to 27\% of the average magnitude of each curve was applied to all curves. 
The necessity for this shift may be due the non-ideality of our Fabry-P\'erot cavity (it is not a lossless cavity, the mirrors are not perfectly aligned, etc.).

Application of the shift to $\Pcirc$ allowed perfect cancellation of the $\Delta k$ and frequency shift at the zero slope locations. 
The correction was then applied to the whole dataset using the shifted $\Pcirc$ curves and the value obtained for $\Ctherm$.
The value of $\Ctherm$ derived from this procedure was 
%$-1.67 \frac{\text{N}}{m \times \mu W}$. 
$-1.67~(\text{N}\text{m}^{-1}\mu \text{W}^{-1})$,
which is in very good agreement with the value, 
$\Ctherm = -2.01~(\text{N}\text{m}^{-1}\mu \text{W}^{-1})$,
which is derived from the COMSOL simulation.

\subsection{Estimation of optical force gradient and 
delay time from the observed effective $Q$ and frequency shift}
From the measured resonant frequency shift, $\df$, and effective $Q$ factor, $\Qeff$, we can estimate the magnitude of the optical spring constant, $\kopt$, and its time delay, $\tau$.
We use the theory developed by Metzger \textit{et al.} 
\cite{metzgerOpticalSelfCooling2008}.
For a small resonant frequency shift $\df \ll \fO$, 
\begin{equation}
    \df = -\frac{\fO}{2k_c} \frac{\partial \Fopt}{\partial z}\frac{1}{1+(2\pi f_0 \tau)^2} 
    \label{eq:deltaf}
\end{equation}
\begin{equation}
    \Delta \gamma \equiv \geff - \gO = \frac{2\pi f_0}{k_c}
    \frac{\partial \Fopt}{\partial z} \frac{2\pi f_0 \tau}{1+ (2\pi f_0 \tau)^2}
    \label{eq:deltag}
\end{equation}
where $\fO$, $\kc$, and $\gO$  are the resonant frequency, the spring constant, and the intrinsic damping constant of the cantilever, respectively.
%and are the effective and s, respectively.
$\tau$ is a time constant which describes the delayed response of the optical force with respect to the cantilever deflection.
The derivation of Eqs.~\ref{eq:deltaf} and \ref{eq:deltag} can be found in Supporting Information~S3.
Note that the damping due to the optical force, $\Delta \gamma$, can be obtained from the experiments as 
\begin{equation}
 \Delta \gamma = \geff - \gO 
 = \frac{\omega_\text{eff}}{\Qeff} - \frac{\omega_0}{\QO} 
 \approx \omega_0 \left(\frac{1}{\Qeff}  - \frac{1}{\QO} \right)
\end{equation}
where $\omega_\text{eff} = 2\pi (\fO + \df)$.
Taking the ratio of Eq.~\ref{eq:deltag}  to Eq.~\ref{eq:deltaf}
yields an expression for the time constant,
\begin{equation}
    \tau = -\frac{1}{8\pi^2 \fO}\frac{\Delta \gamma }{\df}
\end{equation}
We can obtain $\tau$ from the experimentally observed $\df$ and $\Delta \gamma$.
Once we obtain $\tau$, we can estimate the optical spring constant, $\kopt$ from
\begin{equation}
    \kopt = \frac{\partial \Fopt}{\partial z} = -2\kc \frac{\df}{\fO} \{1 + (2\pi \fO \tau)^2 \}
    \label{eq:kopt_from_deltaf}
\end{equation}
or
\begin{equation}
     \kopt = \frac{\partial \Fopt}{\partial z} 
     = k_c \frac{\Delta \gamma}{(2\pi f_0)^2}  
     \frac{1+ (2\pi f_0 \tau)^2}{ \tau}
   \label{eq:kopt_from_deltag}
\end{equation}
We used the corrected values of the frequency shift to calculate $\tau$ and $\kopt$. 
At each measurement point, we used the damping and $\fO$ to derive $\Delta \gamma$ and $\df$, taking the values at the lowest laser power as the $\gO$ and $\fO$. 
We also used the fitted circulating power curves to calculate the expected spring constant due solely to radiation pressure. 

\begin{figure}[h]
\centering
\includegraphics[width=.45\textwidth]{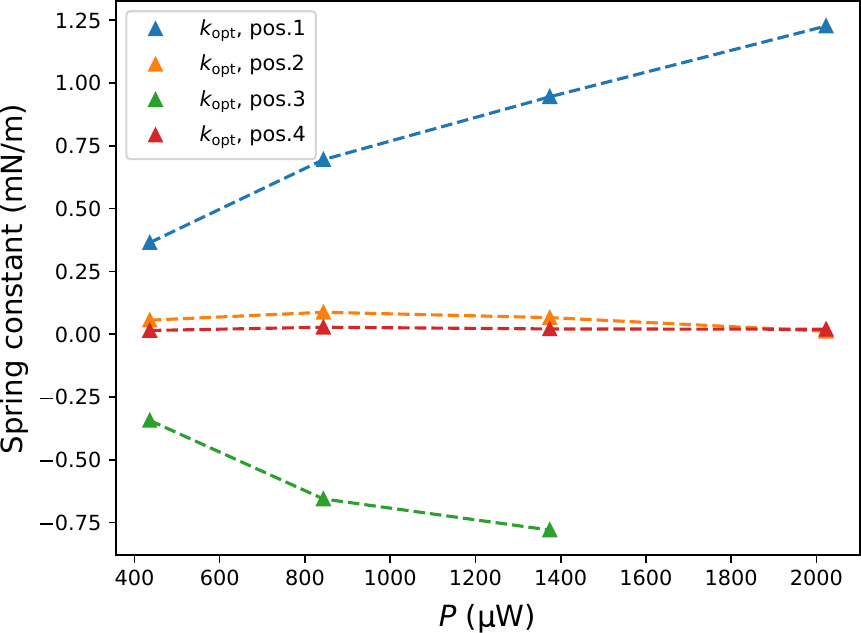}
\caption{\label{fig:k_opt} Estimated $\kopt$ vs average excitation laser power at the positions 1-4.}
\end{figure}

Figure~\ref{fig:k_opt} shows the calculated optomechanical spring constant from the above equations.
The radiation pressure comprises a small fraction (roughly $4\sim5$\%) of the total spring constant (see Supporting figure Fig.~S7, implying that the effect is dominated by photothermal force, which is consistent with other studies \cite{metzgerOpticalSelfCooling2008, holscherEffectiveQualityFactor2009a, miyaharaOpticalExcitationAtomic2020d}.
We obtained the delay time, $\tau \approx 1~\mu$s, from the data at position 1, which corresponds to the phase delay of $\sim 60$~degree.
This is consistent with the above observation 
because the photo-decay time (which is on the order of fractions of a picosecond)  of our low finesse FP cavity is much shorter than $\tau$ we obtained.

\section{Conclusion}
We demonstrated a simple and effective technique to control the effective $Q$ factor of an AFM cantilever using cavity optomechanical force. 
The two-laser technique offers the ability to alter the effective $Q$ factor without changing the detection sensitivity. 
Optomechanically controlled damping offers new options to AFM users. 
For instance, one can operate the cantilever at high effective $Q$ factor for frequency modulation AFM, and then reduce the effective $Q$ factor to help facilitate $Q$ control tapping mode even in vacuum. 
Capability of adjusting the effective $Q$ factor will be useful to optimize the dynamic response and noise performance of phase-locked loop used for frequency modulation mode AFM \cite{demirUnderstandingFundamentalTradeoffs2021a}.
The influence of optical absorption on the cantilever's temperature and resonant frequency was modeled and simulated, providing a more complete picture of the interactions involved in interferometric sensing. Our finite element simulation provided a temperature vs absorbed power relation. For an absorbed power of $1600~\mu$W, which is a plausible value for our interferometer, a significant temperature rise of $\sim 2.1$~K was observed. It is useful to understand the effect of laser power in interferometric sensing on the cantilever temperature, especially in low temperature atomic force microscopy.

\section{Acknowledgment}
We acknowledge the support from National Science Foundation under grants 
NSF CAREER (grant No. DMR2044920) and the NSF Major Research Instrumentation program (grant No. DMR-2117438) and Partnership for Research and Education in Materials via the Division for Materials Research (DMR; award No. DMR-2122041).
We also acknowledge the support from Texas State University.

\bibliographystyle{aipnum4-1}

\bibliography{Q-control}

% \documentclass[aps,pre,showpacs,superscriptaddress,groupedaddress,preprint]{revtex4-2}
% %\documentclass[main.tex]{subfiles}
% \usepackage{graphicx}
% \usepackage{amsmath}
% %\usepackage{hyperref}
% %\usepackage{lineno}
% \usepackage{physics}
% \usepackage{float}
% %\usepackage{caption}
% \usepackage{newtx}
% \usepackage{hyperref}
% %\linenumbers

% *****************************************************************************************************************************************************
% KHB EDITS START HERE

\setcounter{equation}{0}
\setcounter{figure}{0}
\setcounter{section}{0}

\renewcommand{\theequation}{S\arabic{equation}}
\renewcommand{\thefigure}{S\arabic{figure}}

\renewcommand{\thesection}{S\arabic{section}}
\renewcommand{\thesubsection}{\thesection.\arabic{subsection}}
\renewcommand{\thesubsubsection}{\thesubsection.\arabic{subsubsection}}

\renewcommand{\bibnumfmt}[1]{[S#1]}
\renewcommand{\citenumfont}[1]{S#1}

\onecolumngrid

\newpage
\pagebreak

\begin{center}
\textbf{\large Supporting information for\\ \vspace{2mm}
   Control of quality factor of atomic force microscopy cantilever by cavity optomechanical effect}

\vspace{5mm}

Noah Austin-Bingamon$^{1}$, Binod D.C.$^1$ and Yoichi Miyahara$^{1,2}$ \\
\normalsize{$^1$Materials Science, Engineering and Commercialization program,
Texas State University, San Marcos, Texas, 78640, USA}\\
\normalsize{$^2$Department of Physics, Texas State University, San Marcos, Texas, 78640, USA}

\end{center}
\makeatletter

\section{Supporting figures for temperature increase of the cantilever due to optical absorption  \label{sec:canti_temperature}}
\begin{figure}[h]
\centering
\includegraphics[width=.8\linewidth]{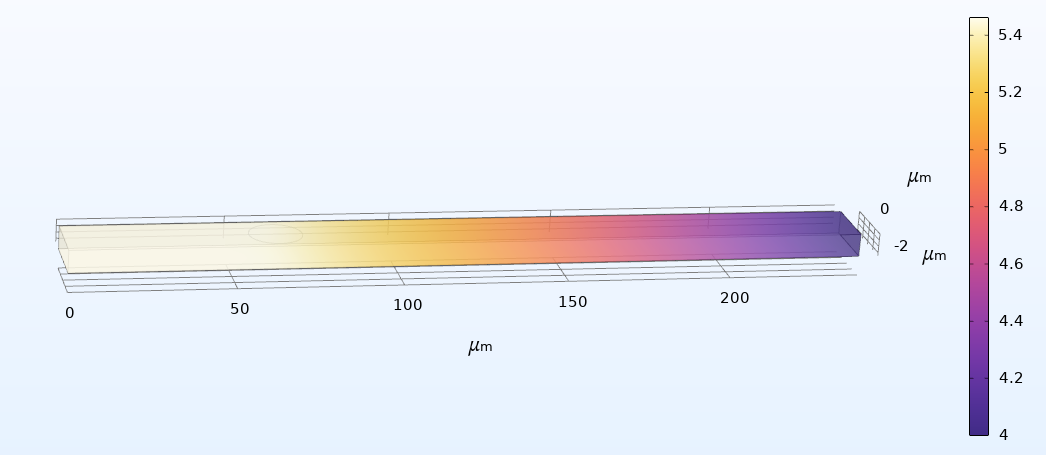}
\caption{\label{fig:cantilever heated comsol} Temperature profile of the simulated cantilever using the highest laser power from the experiment (2022~$\mu$W for 1310~nm and 670~$\mu$W for 1550~nm at fiber end). 
Heat flows into the cantilever in the circular region near the 50~$\mu$m marker. Heat flux is 539~$\mu$W due to the 20~\% absorbed fraction of incident light. The base is fixed at 4~K. }
\end{figure}

\begin{figure}[H]
\centering
\includegraphics[width=.7\linewidth]{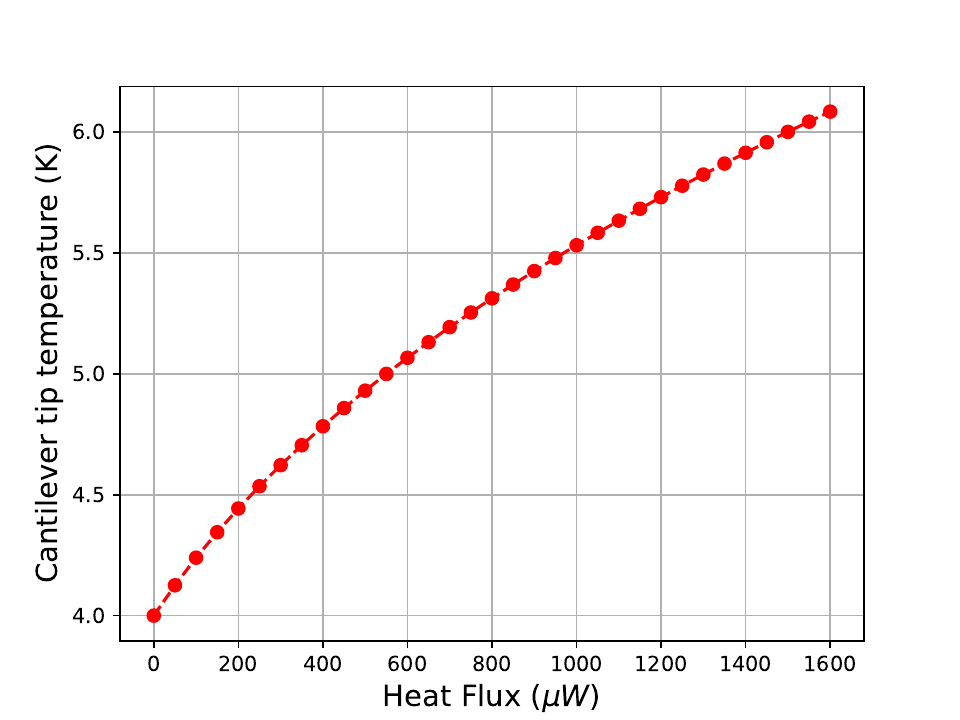}
\caption{\label{fig:cantilever temp comsol} Temperature at the tip of the simulated cantilever for heat flux in range 0-1600 $\mu W$ }
\end{figure}

\section{Fitting of interferograms
\label{sec:fitting}}
\begin{figure}[h]
\centering
\includegraphics[width=.7\linewidth]{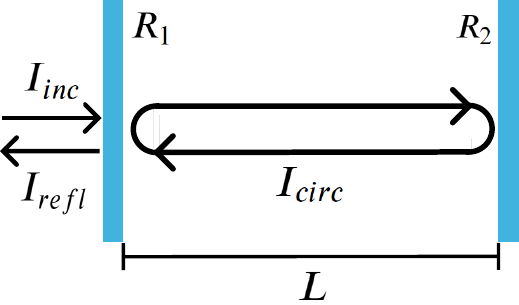}
\caption{\label{fig:FPI diagram} Diagram of Fabry-P\'erot interferometer.}
\end{figure}

We use the Fabry-P\'erot interferometer model reported in Ref.~\cite{Ismail2016} 
for the following analysis. The relevant parameters are illustrated in Fig.~\ref{fig:FPI diagram}

The ratio of the intracavity intensity, $I_\mathrm{circ}$,
to the intensity incident upon mirror 1 , $I_\mathrm{inc}$,
(enhancement factor, $\Acirc'$) is given by
Airy distribution,

\begin{equation}
  \label{eq:A'circ}
  \Acirc ' = \frac{\Icirc}{\Iinc} = \frac{1-R_1}{(1-\sqrt{R_1 R_2})^2 + 4\sqrt{R_1 R_2} \sin^2 \phi}.
\end{equation}
where $\phi$ is the single-pass phase shift between the mirrors
and can be expressed as
\begin{equation}
  \label{eq:phi}
  \phi = \frac{2 \pi d}{\lambda} %\left( z \cos{\theta_t + 2 \delta} \right)
\end{equation}
where $d$ is the distance between the two mirrors (cavity length) and $\lambda$ is wavelength.
Similarly, the enhancement factor for
the total back reflected light with respect to $I_\mathrm{inc}$
is given by

\begin{equation}
  \label{eq:reflectance}
  \frac{\Irefl}{\Iinc} = \frac{(\sqrt{R_1}-\sqrt{R_2})^2 + 4\sqrt{R_1 R_2} \sin^2\phi}
  {(1-\sqrt{R_1 R_2})^2 + 4\sqrt{R_1 R_2} \sin^2 \phi}.
  %  = \frac{F (\alpha  +  \sin^2{\phi})}
  %  {1 + F \sin^2{\phi}}
\end{equation}

\begin{figure}[h]
\centering
\includegraphics[width=1\linewidth]{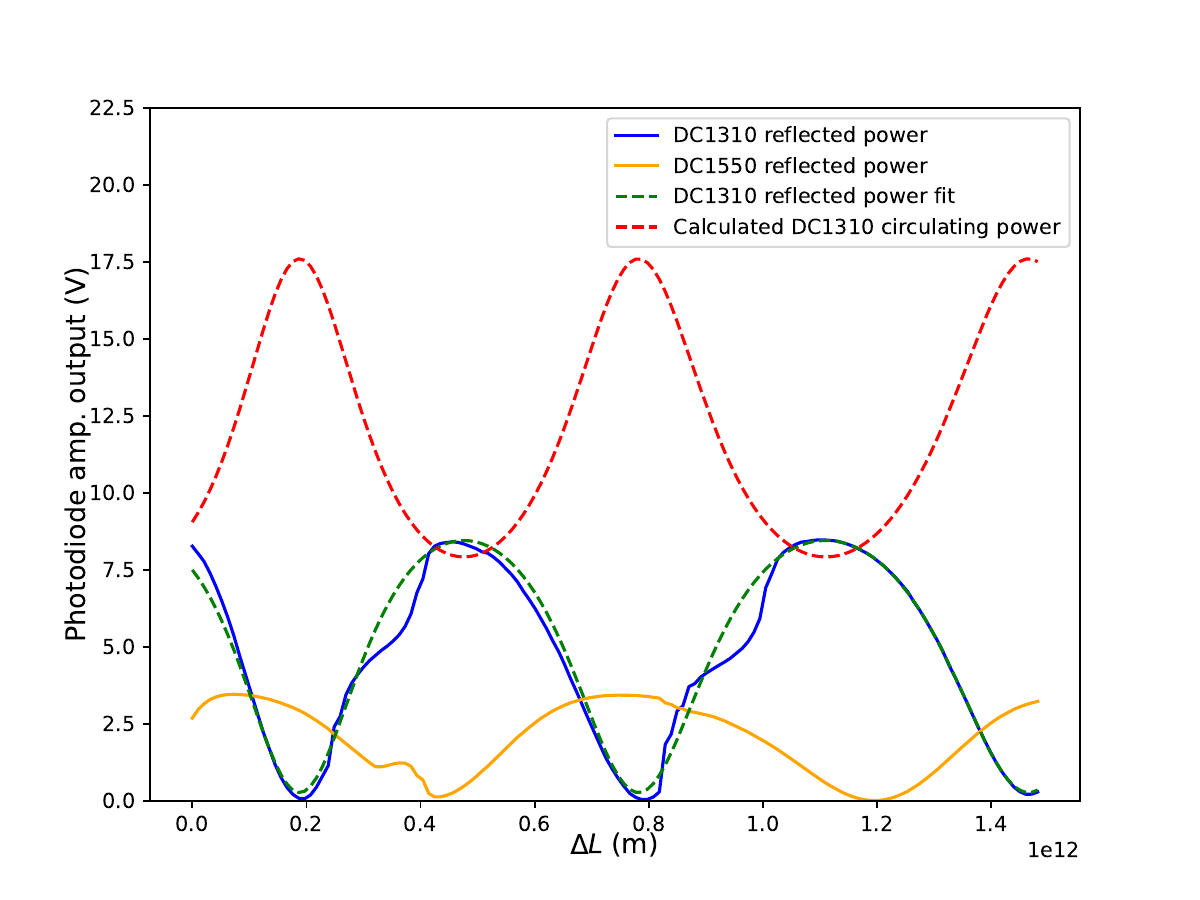}
\caption{\label{fig:1310 fitting} Fitting of the interferogram for the 1310 nm laser according to Eq.~\ref{eq:reflectance}. The predicted circulating power curve is also shown according to Eq.~\ref{eq:A'circ}. The curves are shown in units of volts at the photodiode amplifier output, which is proportional to the power. The scaling of the curve has no effect on the fitted cavity parameters $R_1$ and $R_2$, so no calibration was performed. Fitted parameters are $\lambda$ = 1324 nm, $R_1$ = 0.251, $R_2$ = 0.155, $I_\text{inc}$ = 15.158 (note that $I_\text{inc}$ is not calibrated to the true power here).}
\end{figure}

The fitting, done by SciPy's optimize function, was performed for the 1310 nm reflection signal over a range of 2.5 fringes using Eq.~\ref{eq:reflectance}. Fitting results are shown in Fig.~\ref{fig:1310 fitting}.

\begin{figure}[h]
\centering
\includegraphics[width=1\linewidth]{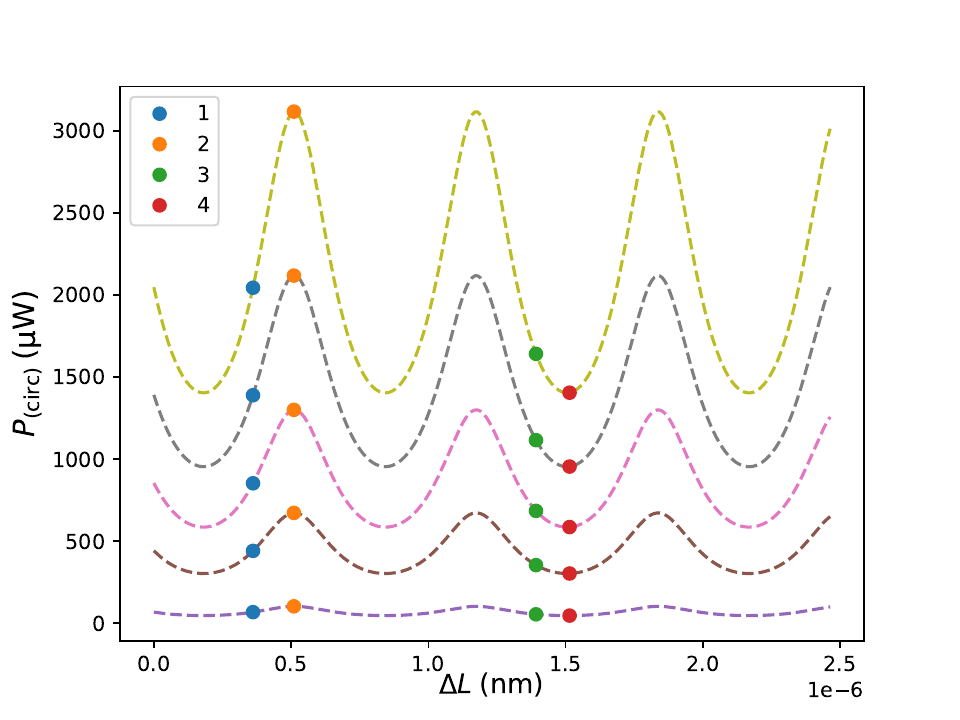}
\caption{\label{fig:circ power from fit} Circulating power for the 1310 nm laser derived from the fitted cavity parameters $R_1$ and $R_2$ and using the values of $I_\text{inc}$ from the experiment. Each curve corresponds to one laser power used. 
See Eq.~\ref{eq:A'circ}. $I_\text{inc}$ was calculated from the power at the free fiber end using the fitted reflectance $R_1$ according to $I_\text{inc} = I_\text{laun}/(1-R_1)$.}
\end{figure}

The key parameters are the reflectivities of the two cavity mirrors, as these allow calculation of the circulating power in the cavity for a given $I_\text{inc}$ and cavity length. Circulating power curves were calculated from the fitted parameters and the laser powers used in the experiment and are shown in Fig.~\ref{fig:circ power from fit}. The measurement positions, as shown in Fig.~2 in the body text, are indicated by the colored dots.

\clearpage

\section{Derivation of equations 1 and 2  \label{sec:derivation}}
Adapting the theory by Metzger \textit{et al.} \cite{Metzger2008}, 
the effective spring constant, $\keff \equiv m \omega_\text{eff}^2$, can be described as follows:
\begin{equation*}
    \keff = \kc \left( 1 - \frac{1}{1+\omega^2 \tau^2}  \frac{\partial \Fopt}{\partial z}\frac{1}{\kc}\right)
   \end{equation*}
\begin{equation}
    \frac{\keff - \kc}{\kc} 
    = \frac{\weff^2 - \wO^2}{\wO^2}
    =- \frac{1}{1+\omega^2 \tau^2}  \frac{\partial \Fopt}{\partial z}\frac{1}{\kc}
    \label{eq:keffs} 
\end{equation}
Considering a small (angular) resonant frequency shift, $\dw \equiv \weff - \wO$, 
$\abs{\dw} \ll \wO$,
\begin{equation}
    \frac{\weff^2 - \wO^2}{\wO^2} 
    =\frac{(\wO + \dw)^2-\wO^2}{\wO^2}  
    \approx 2\frac{\dw}{\wO} = 2 \frac{\df}{\fO}
    \label{eq:df_approx}
\end{equation}
From Eqs. \ref{eq:keffs} and \ref{eq:df_approx}, 
we obtain Eq.~1 in the body text.
\begin{equation}
    \df = -\frac{\fO}{2k_c} \frac{\partial \Fopt}{\partial z}\frac{1}{1+(2\pi f_0 \tau)^2} 
    \label{eq:deltafs}
\end{equation}

Similarly, the effective damping constant, $\geff \equiv \weff/\Qeff$, can be expressed by
\begin{equation*}
    \geff = \gO \left( 1 + \QO \frac{\wO \tau }{1+\omega^2 \tau^2} \frac{\partial \Fopt}{\partial z}\frac{1}{\kc}\right)
\end{equation*}
\begin{equation}
    \Delta \gamma = \geff - \gO = \left( \gO \frac{\wO}{\gO} \frac{\wO \tau }{1+\omega^2 \tau^2} \frac{\partial \Fopt}{\partial z}\frac{1}{\kc}\right)
   =\frac{\wO}{\kc} \frac{\wO \tau }{1+\omega^2 \tau^2} \frac{\partial \Fopt}{\partial z}   
    \label{eq:geffs}
\end{equation}

From Eq.~\ref{eq:geffs} and considering $\omega \approx \wO$, we obtain Eq.~2 in the body text.

\begin{figure}[h]
\centering
\includegraphics[width=1\linewidth]{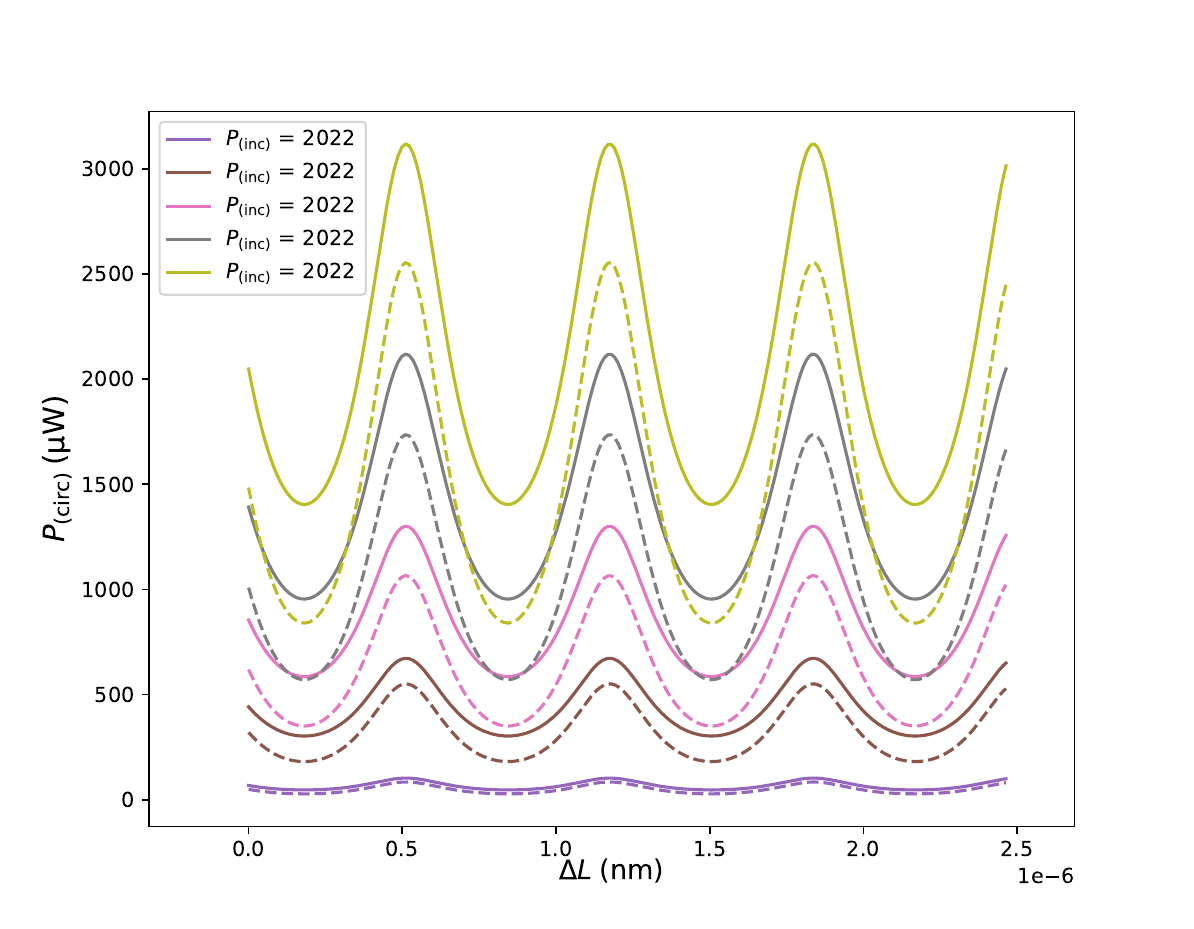}
\caption{\label{fig:shifted circ power} The shift in the circulating power curves needed to adequately cancel the thermal frequency shift at both zero slope positions. Shifted curves are shown as dotted lines. Each curve is shifted down by 27.5\% of its average value.}
\end{figure}

\begin{figure}[h]
\centering
\includegraphics[width=.9\textwidth]{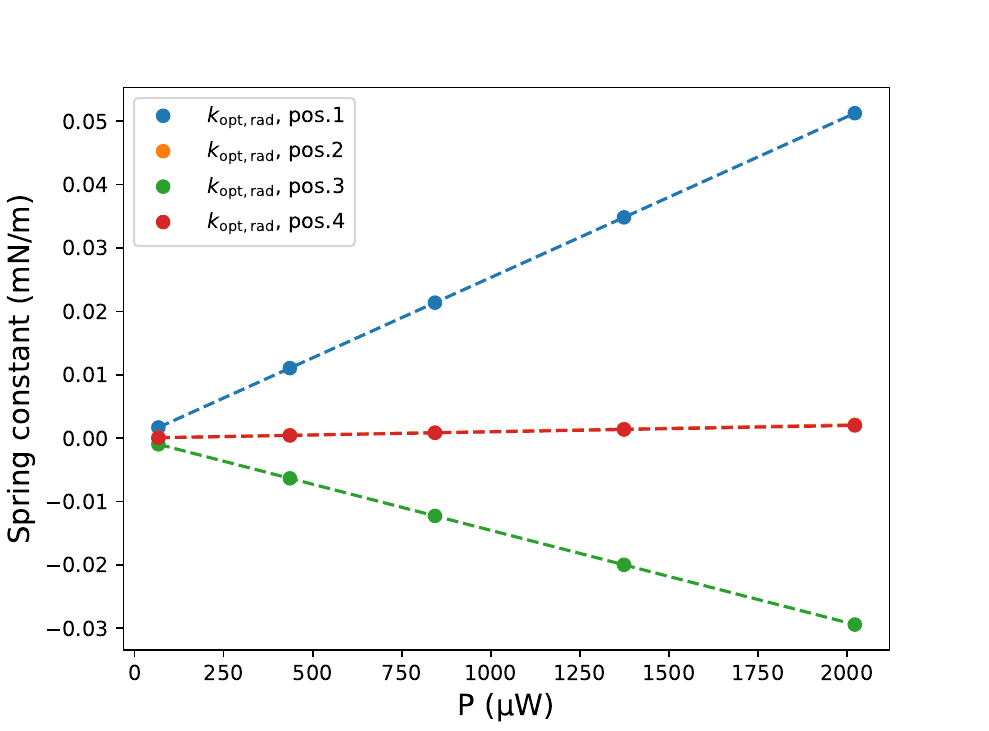}
\caption{\label{fig:k_opt,rad} $\kopt$ due to radiation 
pressure vs average excitation laser power. 
The curve for position 4 overlays that of position 2. Force curves were obtained from the circulating power assuming a perfectly reflective surface as $F_\text{rad}=2 P_\text{inc}/c$, where $P_\text{inc}$ is the incident power and $c$ is the speed of light. Spring constants were derived by taking the numerical derivative at each measurement point.}
\end{figure}

% \bibliographystyle{aipnum4-1}
% \bibliography{Q-control}

%

% \end{document}

\end{document}